\documentclass[a4paper,11pt]{article}
\usepackage[usenames]{color}

\usepackage{amsfonts,amssymb}
\usepackage{theorem}
\usepackage{rotating}

\usepackage{cite}
\usepackage{epsfig}

\newcommand{\comment}[1]{}

\begin{document}
%
%
\Large
{\bf
\centerline{On the Spectrum of Lattice Massive $SU(2)$ Yang-Mills}}

\large
\rm

\begin{center}
Ruggero~Ferrari$^{a,b,}$\footnote{e-mail: {\tt ruggero.ferrari@mi.infn.it}}
\end{center}

\small
\medskip
\begin{center}
$^a$INFN, Sezione di Milano\\
via Celoria 16, I-20133 Milano, Italy\\
and\\
$^b$
Center for Theoretical Physics\\
Laboratory for Nuclear Science
and Department of Physics\\
Massachusetts Institute of Technology\\
Cambridge, Massachusetts 02139\\
(MIT-CTP 4477, June 2013 )
\end{center}

\normalsize

%
\normalsize

\begin{quotation}
\rm{\Large Abstract:}
On the basis of extended simulations we provide some
results concerning the spectrum of Massive  $SU(2)$ Yang-Mills 
on the lattice. 
We study the ``time'' correlator of local gauge invariant
operators integrated over the remaining three dimensions. The
energy gaps are measured in the isospin  $I=0,1$
and internal spin $J=0,1$ channels. 
\par 
No correlation is found in the
$I=1,J=0$ channel. 
In the $I=1, J=1$ channel and far from the critical mass value $m_c$
the energy gap roughly  follows the bare value $m$ (vector mesons). 
In approaching the  critical value  $m_c$ at $\beta$ fixed, 
there is a bifurcation of the energy gap:
one branch follows the value $m$, while the new is much larger
and it shows a more and more dominant weight.
This phenomenon might be the sign of two important
features: the long range correlation near the fixed
point at $\beta \to \infty$ implied by the low energy gap
and the screening (or confining) mechanisms across the $m=m_c$
associated to the larger gap.

The $I=0, J=0,1$ gaps are of the same order of magnitude, typically
larger than the $I=1, J=1$ gap (for $m>>m_c$).
For $m\sim m_c$ both $I=0$ gaps have a dramatic drop 
with minima  near the value $m$. This behavior might correspond
to the formation of $I=0$ bound states both in the 
$J=0$ and $J=1$ channels.
\end{quotation}
\newpage
\normalsize
\rm

\section{Introduction}
\label{sec:intro}
%
In the present paper we continue the study of the Massive $SU(2)$
Yang-Mills (MYM) theory on the lattice, initiated in Ref.
\cite{Ferrari:2011aa} 
and further pursued in Ref. \cite{Bettinelli:2012qk}. 
Let us remind why we consider the model of great interest.
Recently a Massive Yang-Mills theory for $SU(2)$
has been formulated in the continuum 
in Refs. \cite{Bettinelli:2007tq},\cite{Bettinelli:2007cy}.
The  mass  is introduced {\sl \`a la } St\"uckelberg.
Since the theory is nonrenormalizable a new subtraction
strategy is necessary. The strategy has been developed in
a series of papers (\cite{Ferrari:2005ii},\cite{Ferrari:2005va},
\cite{Bettinelli:2007zn}) and it is based on a Local Functional
Equation (LFE) for the vertex functional and on dimensional
subtraction. Although the subtraction procedure
has been successfully applied to massless\cite{Ferrari:2005ii}  
and massive  \cite{Ferrari:2010ge} nonlinear sigma model, to
the low energy electroweak model
\cite{Bettinelli:2008ey},
\cite{Bettinelli:2008qn},
\cite{Bettinelli:2009wu} and to field-coordinate transformations
\cite{Ferrari:2009uj}, still nonrenormalizability has 
unpleasant consequences
for the high energy behavior in most of the listed cases
(unitarity violations). It has
been suggested that such nonrenormalizable theories, once made
finite by the appropriate subtraction strategy, undergo to
a phase transition \cite{Ferrari:2012kd}, \cite{Ferrari:2011bx}
at very large energies. 
This conjecture might
be investigated in a nonperturbative approach, as in a lattice
model. This is the rational for considering a Massive Yang-Mills
lattice gauge theory: the model has the same local gauge symmetry as in the
continuum and one has the possibility to avoid completely any
gauge fixing. The challenge consists in comparing the lattice 
and the continuum amplitudes, in mapping the parameters
and in evaluating the limit of validity of the lattice model 
as a phenomenological theory. In \cite{Ferrari:2011aa} 
the existence of a Transition Line (TL)  $m=m_{_{\rm TL}}(\beta)$ 
in the  $(\beta,m^2)$ space has been confirmed.
Along this line, from the end point $\beta_e \sim 2.2$ 
through $\beta \to \infty$,
both energy and order parameter have a very steep inflection,
whose derivative increases with the lattice size
(as discussed later in Section  \ref{sec:model}). The line 
separates  the deconfined phase from a supposedly confined phase.
For $\beta< \beta_e$ the transition through the line  is smooth.
Thus we denote the TL by $m=m_c(\beta)$ for $\beta> \beta_e$.
\par
In Ref. \cite{Bettinelli:2012qk} we have compared global quantities
as energy and order parameter evaluated by Monte Carlo in the
lattice and two-loop calculations in the continuum. The results are
suggestive of a good agreement.
\par
The present paper is devoted to the investigation of the particle
content of the Massive Yang-Mills on the lattice in the deconfined
region of the parameter space ($\beta,m^2$). We look at the 
energy gap in the time correlator of suitable gauge invariant
operators mediated over the other dimensions (zero three-momentum
states). These operators are easily associated to particles
with isospin $I=0,1$ and spin $J=0,1$ in the deconfined
phase. The fit is done with the function
\begin{eqnarray}&&
g(x) = \frac{1}{2}(f(x)+f(L-x))
\nonumber\\&&
f(x) = b_1 e^{-\Delta_1 x} + b_2 e^{-\Delta_2 x},
\label{intro.1a}
\end{eqnarray}
where $L$ is the size of the lattice (an integer)
and $L^4$ gives the number of the lattice sites.
We considered mainly lattices of size $24^4$ and the measures
are performed on $10^4$ configurations each separated by 15 updatings.
Statistical errors are evaluated by using bins of size 100.
\par
This excruciating analysis  is limited to few values 
of $\beta=1.5,3,10,40$, being the end-point $\beta_e\sim 2.2$.
We find no correlation in the channel $I=1,J=0$; i.e. no scalars with
flavor 1. In the channel $I=1,J=1$ (gauge vector mesons) the value of
the gap is close to the bare mass $m$ for $m\ge 1$.
For $m$ near $m_c$ (thus $\beta =1.5$ excluded) a bifurcation occurs, 
i.e. the fit
requires two energy gap parameters. The lower follows the $m$ value
while the second is much larger. {\sl A posteriori} this bifurcation
looks necessary if we expect a large correlation length
near $\beta \to \infty$ ($ \Longrightarrow m_c \to 0$)
and a larger gap for the establishment of confinement across the TL.
\par
A similar situation is present in the channels $I=0,J=0,1$;
however the bifurcation sets on nearer the TL than in the
vector mesons channel.  The numerical values in general 
follow the pattern $\Delta_{I=0J=1}\simeq\Delta_{I=0J=0}
>>m$. For $m\sim m_c$ the lower energy gap in the $I=0$
channels drops to values of the order of $m$. In this
region the larger gap in the $I=1,J=1$ channel is
dominant; therefore long living resonant states might
develop in the $I=0, J=0,1$ channels.
\par
This scenario of the spectrum open many interesting questions.
We mention here a couple of them.
The onset of confinement across the TL for $\beta>> \beta_e$
is clearly related to the bifurcation of
gap when $m \to m_c$. The present work spots the
point where it is possible to study  the mechanism of
confinement at its onset. 
A further question of great interest is whether 
a bound state of two vector mesons exists near
the TL, i.e. where the correlation length becomes
larger. We shall illustrate this phenomenon 
with some pictures later on.
\par
The lattice model is of great interest by itself: the phase diagram in
the parameter space $(\beta,m^2)$ is very intriguing. 
The TL at large $\beta$ is compatible with
$\beta m^2\sim 0.64$, i.e. the TL points to the critical
point of the $O(4)$ nonlinear sigma model \cite{Baaquie:1992cw}. 
\par
The same lattice gauge model 
has been studied previously (see 
\cite{Fradkin:1978dv}-\cite{Bonati:2009pf}) as an example
of Higgs mechanism with a frozen length. We agree on
the position of the TL, but we have no definite results
on the exact nature of the phase transition, beyond the
presence of a steep inflection which becomes more and
more strong by increasing $\beta$.
\par
Further work is necessary in order to establish the
character of the phase transition across the TL. Moreover
it is very important to interpret the model in the
limit of $\beta \to \infty$  where a fixed point
is expected \cite{Berg:1988cy}. In the limit some correlation
length should become very large.
\par
The relation between the lattice model and the continuum
theory is not discussed here. We postpone this complex topic
to a future work.


\section{The Model}
\label{sec:model}
The present Section is devoted to the recollection of
the essentials of the model. More details are
given in Refs. \cite{Ferrari:2011aa} and \cite{Bettinelli:2012qk}.
\par
The action on the cubic lattice of size $N\equiv L^4$ with  sites $x$
and links $\mu$ is
\begin{eqnarray}
S_{ L}=\!\! \frac{\beta}{2} \,\, {\mathfrak Re} \sum_\Box Tr\big\{1-  U_\Box\big\}
+ \frac{\beta}{2} m^2 {\mathfrak Re}
\sum_{x\mu}Tr\Bigl\{1- \Omega(x)^\dagger U(x,\mu)\Omega(x+\mu)\Bigr\},
\label{lat.2}
\end{eqnarray}
where the sum over the plaquette is the Wilson action \cite{Wilson:1974sk}.
The link variables $U(x,\mu)$ and the site variables $\Omega(x)$ are
elements of the $SU(2)$ group.
\par
The action is invariant under the  {\bf local-left} transformations 
$g_{\scriptstyle{\scriptstyle{L}}}(x)\in SU(2)_L$ 
and the {\bf global-right }   transformations $g_{\scriptstyle{R}}\in SU(2)_R$
\begin{eqnarray}
\!
\scriptstyle{SU(2)_L}\left\{
\begin{array}{l}
\Omega'(x) = g_L(x)\Omega(x) \\
U'(x,\mu) = g_L(x) U(x,\mu)  g^\dagger_L(x+\mu)
\end{array} \right. \!\!, ~\,\,
\scriptstyle{SU(2)_R}\left\{
\begin{array}{l}\Omega'(x) = \Omega(x)g_R^\dagger
\\  U'(x,\mu) = U(x,\mu)
\end{array} \right. \!.
\label{stck.5}
\end{eqnarray}
We would like to stress the importance of this invariance
property, in particular because in the nonrenormalizable 
continuum Minkowskean
theory it is the starting point for the removal of the
ultraviolet divergences of the  loop expansion. In fact
the invariance of the path integral measure ensures the
validity of the LFE for the generating functionals (e.g.
the vertex functional) \cite{Ferrari:2005ii}.
\par
\begin{figure}
\epsfxsize=100mm
\centerline{\epsffile{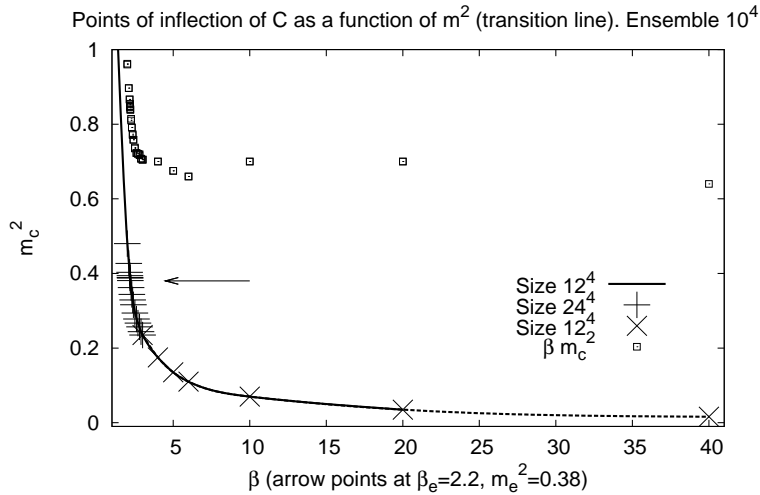}}
\caption{The transition line. The arrow marks the position of the end
point. {In the figure data from previous analysis have been used and 
the statistical errors are not displayed
since they are too small to be shown.}}
\label{fig.1}
\end{figure}

The quantity $(D=4)$
\begin{eqnarray}
{\mathfrak C} := \frac{1}{2D N}\Big\langle
\sum_{x\mu}Tr\Bigl\{\Omega(x)^\dagger U(x,\mu)\Omega(x+\mu)\Bigr\}
\Big\rangle
\label{model}
\end{eqnarray}
is taken as order parameter. It has the symmetry property
${\mathfrak C}(\beta,-m^2) = {\mathfrak C}(\beta,m^2)$.
The TL is given by the {\sl loci}
in the $(\beta,m^2)$ plan
where ${\mathfrak C}$ has an inflection as function of $m^2$
for given $\beta$ as shown in Fig. \ref{fig.1}.
\par 
For large $\beta$ the TL approaches
the critical coupling $\beta m^2 \sim 0.64$ of the $SU(2)$ 
nonlinear sigma model
in 4 dimensions \cite{Baaquie:1992cw}.  Moreover in the
region ${\mathfrak C}\sim 0$ the global-right $SU(2)_R$
charges are screened (or confined), while for ${\mathfrak C}\sim 
\pm 1$ global-right $SU(2)_R$ is unitarely implemented and
vector mesons exist.
\par
The character of the transition across the TL is not yet well
established. The inflection becomes steeper by increasing
the lattice size for $\beta $ larger than
the  end point value: $\beta_e \sim 2.2, m_e^2\sim 0.381$. 
Numerically one cannot easily affirm whether 
it is a first order transition with a small jump or a second
order  or even a crossover. This question is not under
investigation in the present paper.

\section{Gauge Invariant Fields}
\label{sec:field}
In order to investigate the spectrum in the deconfined 
region in the $(\beta,m^2)$ plane we consider the 
field ($\tau_a$ are the Pauli matrices)
\begin{eqnarray}
C(x,\mu) := \Omega^\dagger(x) U(x,\mu)\Omega(x+\mu)= C_0(x ,\mu) 
+ i\tau_a C_a(x,\mu).
\label{field.1}
\end{eqnarray}
By construction
\begin{eqnarray}
C(x,\mu) \in SU(2).
\label{field.2}
\end{eqnarray}
According to the transformations of eq. (\ref{stck.5})   $C(x,\mu)$
is invariant under local-left transformations 
(usually said ``gauge invariant''), while under the global-right
transformations they have  $I=0$ ($C_0$) and  $I=1$ components
($C_a$). One has
\begin{eqnarray}
C_0(x,\mu)^2 + \sum_{a=1,3} C_a(x,\mu)^2 =1.
\label{field.3}
\end{eqnarray}
Then we get
\begin{eqnarray}
|C_0(x,\mu)| \leq 1
\label{field.4}
\end{eqnarray}
and therefore (from eq. (\ref{model}))
\begin{eqnarray}
|{\mathfrak C}| \leq 1.
\label{field.4.1}
\end{eqnarray}
In the deconfined region we expect the global-right
symmetry to be implemented and therefore
\begin{eqnarray}&&
\langle C_a(x,\mu)\rangle =0
\nonumber\\&&
\langle C_a(x,\mu)  C_b(y,\nu)\rangle =0, \qquad if~a\not =b.
\label{field.5}
\end{eqnarray}
Moreover the symmetry over four-dimensional finite
rotations requires
\begin{eqnarray}
\langle C_a(x,\mu)  C_a(y,\nu)\rangle =0, \qquad if~\mu\not =\nu.
\label{field.6}
\end{eqnarray}
The equations (\ref{field.4.1}), (\ref{field.5}) and (\ref{field.6})
are satisfied by the numerical simulations to a reasonable
level of accuracy.

\section{The Numerical Simulation}
\label{sec:sim}
The spectrum is evaluated in the deconfined phase, 
by considering the two-point function of the 
zero-three-momentum operator
\begin{eqnarray}
C_j(t,\mu) = \frac{1}{L^{\frac{3}{2}}}\sum_{x_1,x_2,x_3} 
C_j(x_1,x_2,x_3,x_4,\mu)|_{x_4=t}
,\quad j=0,1,2,3.
\label{sim.1}
\end{eqnarray}
Then we evaluate the connected correlator
\begin{eqnarray}
C_{jj',\mu\nu}(t) = \frac{1}{L} \sum_{t_0=1,L}
\Big\langle C_j(t+t_0,\mu)C_{j'}(t_0,\nu)\Big\rangle_C.
\label{sim.2}
\end{eqnarray}
According to eqs. (\ref{field.5}) and (\ref{field.6})
the correlator is zero unless $j=j'$ and $\mu=\nu$.
The spin one- and zero- amplitudes  $V$ and  $S$ are extracted by
using the relation
\begin{eqnarray}
C_{jj,\mu\nu}(t) = V_{jj}(\delta_{\mu\nu}-\delta_{\mu 4}\delta_{\nu 4})
+ S_{jj}\delta_{\mu 4}\delta_{\nu 4}.
\label{sim.3}
\end{eqnarray}
Very good fit of the data is obtained by using the
function 
\begin{eqnarray}&&
g(t) = \frac{1}{2}(f(t) + f(L-x))
\nonumber\\&&
f(t) = b_1 e^{-\Delta_1 t} + b_2 e^{-\Delta_2 t} .
\label{sim.4}
\end{eqnarray}
Two exponentials  are needed only for $m\simeq m_c$, as we will
illustrate shortly. Otherwise one single exponential is enough for
the fit.
\par
The expectation values are performed on $10^4$ configurations
created by a Heat-Bath Monte Carlo for a lattice of size $24^4$.
A configuration is stored every 15 updating steps. Statistical errors
are evaluated by using bins of 100 measures.
We consider the values $\beta = 1.5,3,10,40$ and $m^2 < 8$. The first
is interesting since it is outside the TL 
(the "end point'' is at $\beta_e\sim 2.2$).  For $\beta=3$ the TL separates  
different phases and the "coupling constant'' 
($g = \sqrt{4/\beta} \sim 1.155$) is large, while for $\beta=40$ we 
are in the region of weak coupling limit ($g\sim 0.316$) and "near'' 
the fixed point $\beta \to \infty$.
\par
\begin{figure}[t]
\begin{center}
\includegraphics[clip,width=1\textwidth,,height=0.4\textheight]
{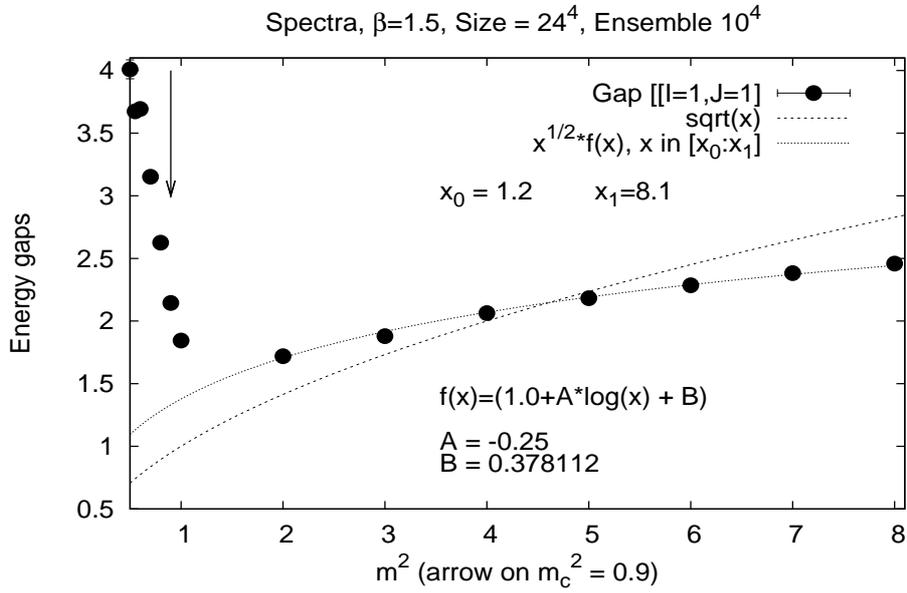}
\end{center}
\caption{{Mass spectrum  of the gauge vector meson 
for $\beta=1.5$.}}
\label{fig.2.0}
\end{figure}
\begin{figure}[t]
\begin{center}
\includegraphics[clip,width=1\textwidth,,height=0.4\textheight]
{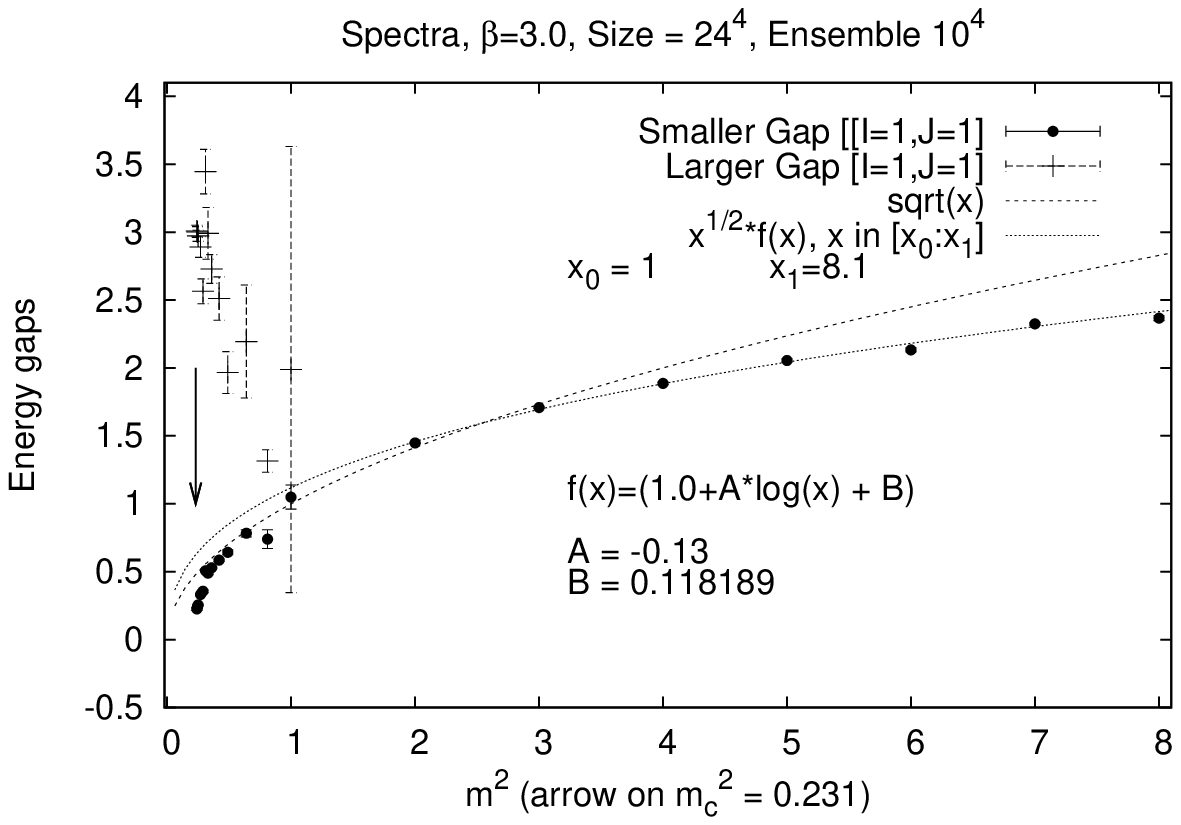}
\end{center}
\caption{Mass spectrum ($m>>m_c$) of the gauge vector meson 
for $\beta=3$.}
\label{fig.2}
\end{figure}
\begin{figure}[h]
\begin{center}
\includegraphics[clip,width=1\textwidth,,height=0.4\textheight]
{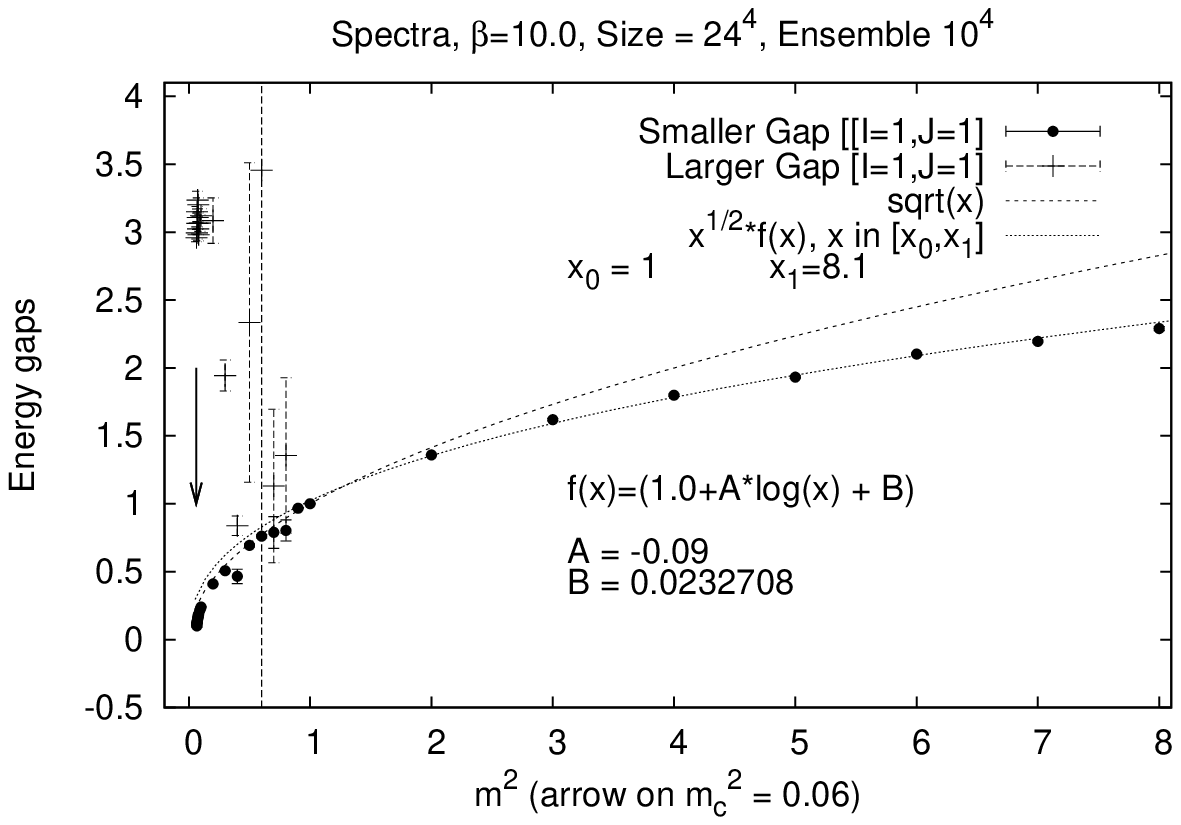}
\end{center}
\caption{Mass spectrum  ($m>>m_c$) of the gauge vector meson 
for $\beta=10$.}
\label{fig.3}
\end{figure}
\begin{figure}[h]
\begin{center}
\includegraphics[clip,width=1\textwidth,,height=0.4\textheight]
{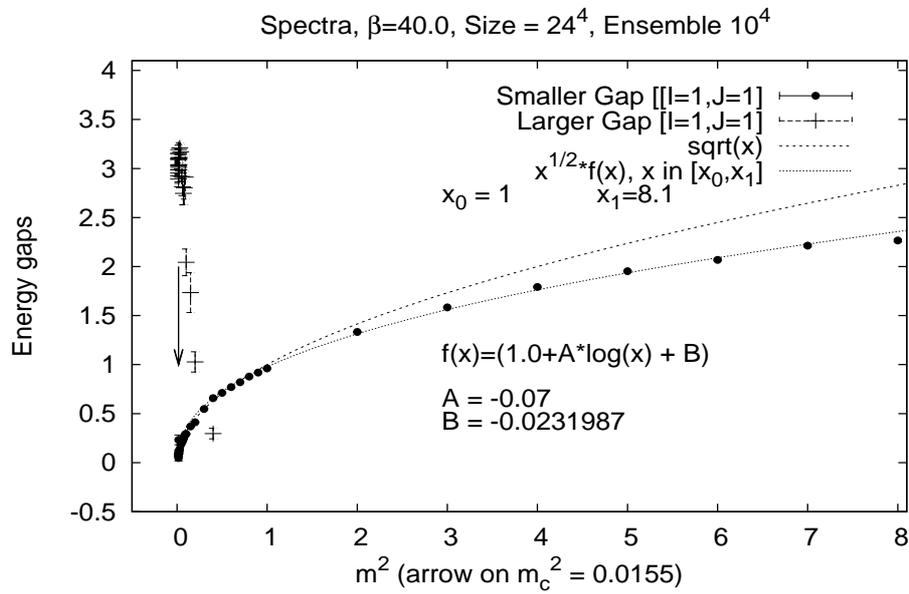}
\end{center}
\caption{Mass spectrum  ($m>>m_c$) of the gauge vector meson 
for $\beta=40$.}
\label{fig.4}
\end{figure}
\begin{figure}[hp]
\begin{center}
\includegraphics[clip,width=1\textwidth,,height=0.4\textheight]
{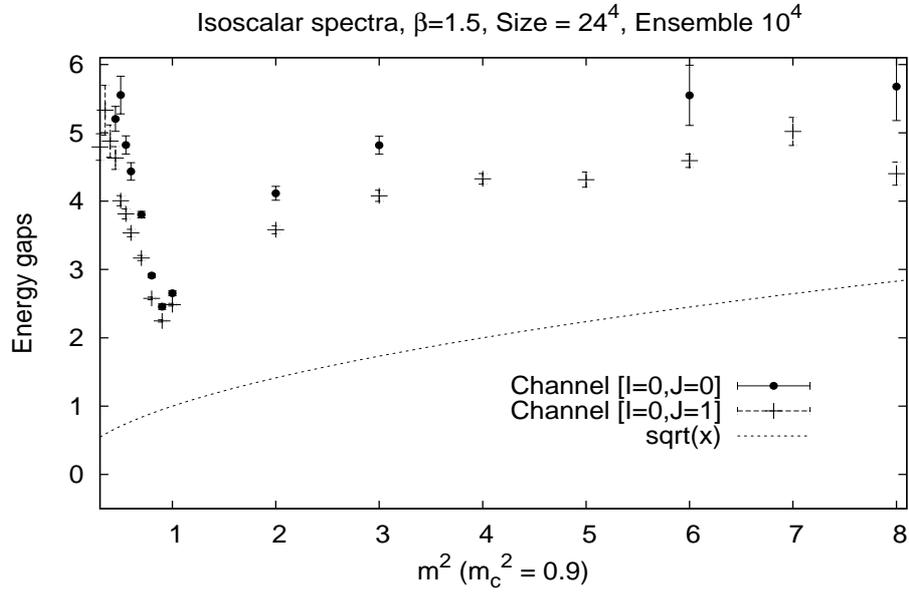}
\end{center}
\caption{Mass spectrum ($m>>m_c$) of the isoscalars
for $\beta=1.5$.}
\label{fig.5.0}
\end{figure}
\begin{figure}[hp]
\begin{center}
\includegraphics[clip,width=1\textwidth,,height=0.4\textheight]
{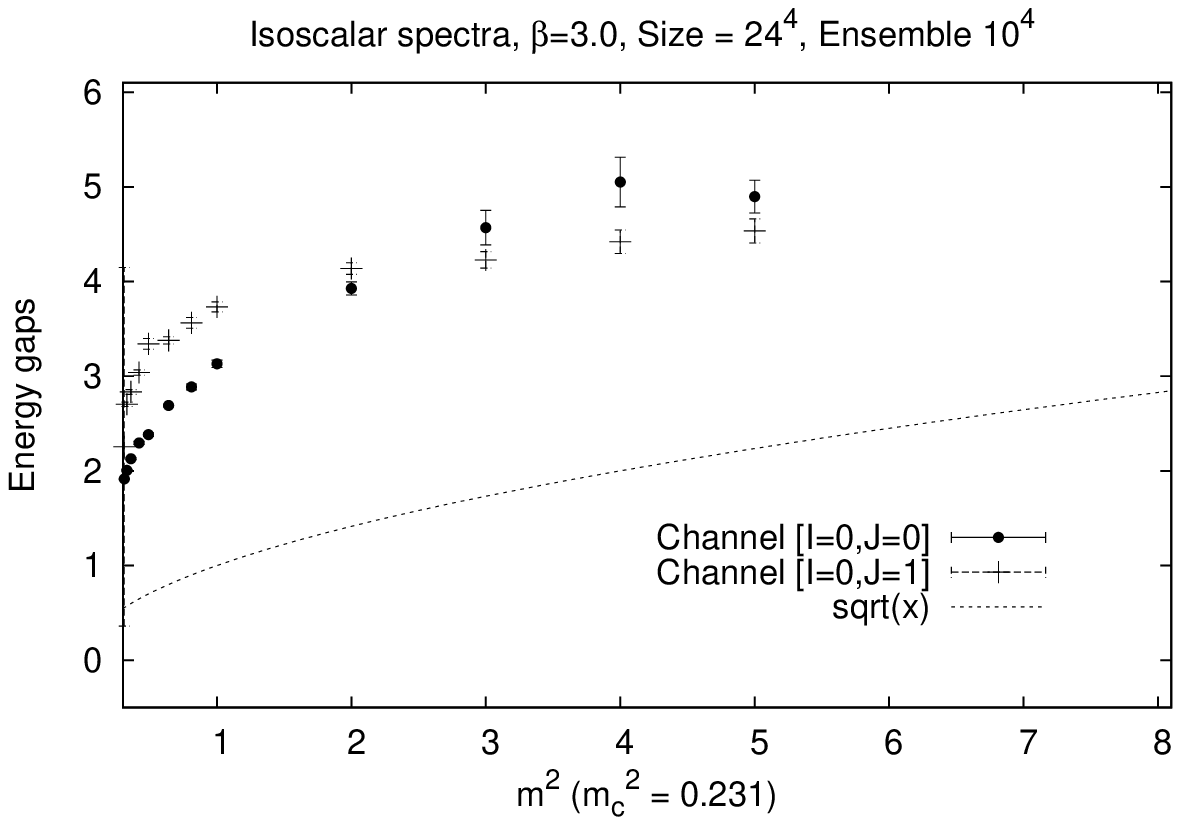}
\end{center}
\caption{Mass spectrum ($m>>m_c$) of the isoscalars
for $\beta=3$.}
\label{fig.5}
\end{figure}
\begin{figure}[hp]
\begin{center}
\includegraphics[clip,width=1\textwidth,,height=0.4\textheight]
{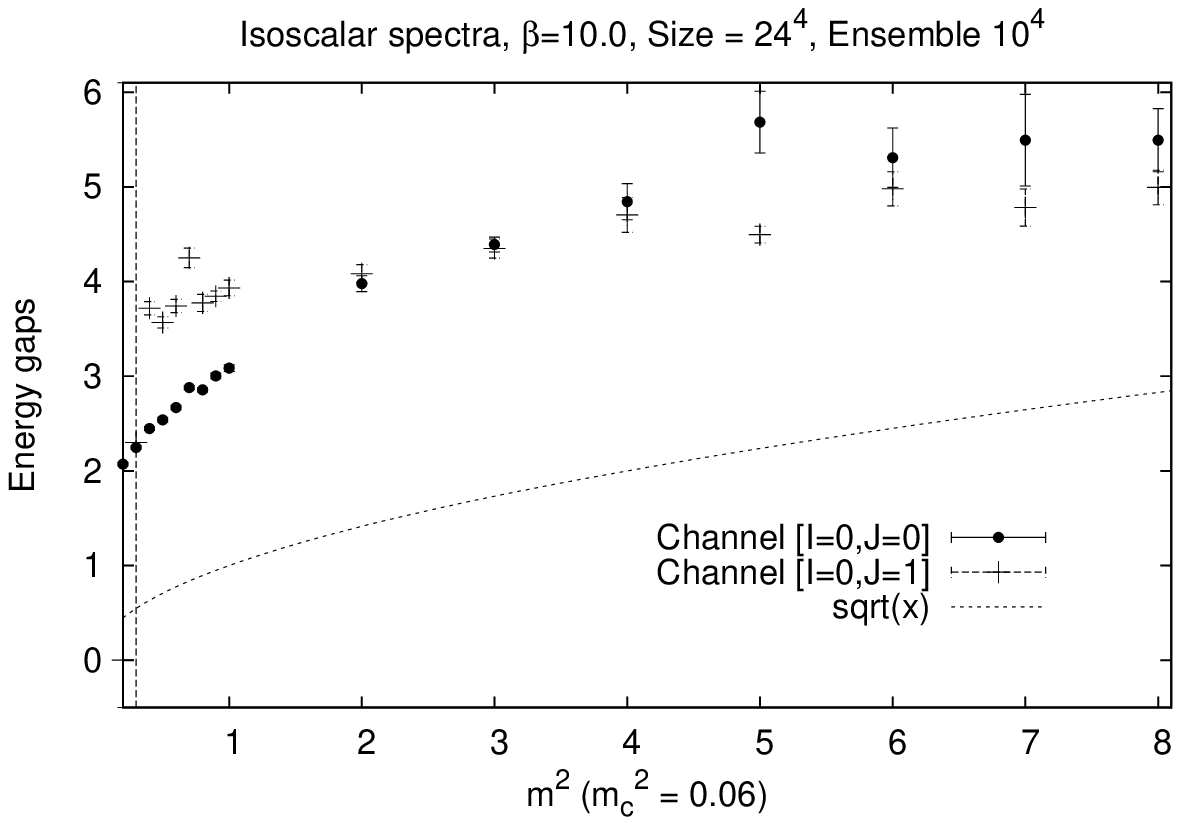}
\end{center}
\caption{Mass spectrum  ($m>>m_c$) of the isoscalars
for $\beta=10$.}
\label{fig.6}
\end{figure}
\begin{figure}[hp]
\begin{center}
\includegraphics[clip,width=1\textwidth,,height=0.4\textheight]
{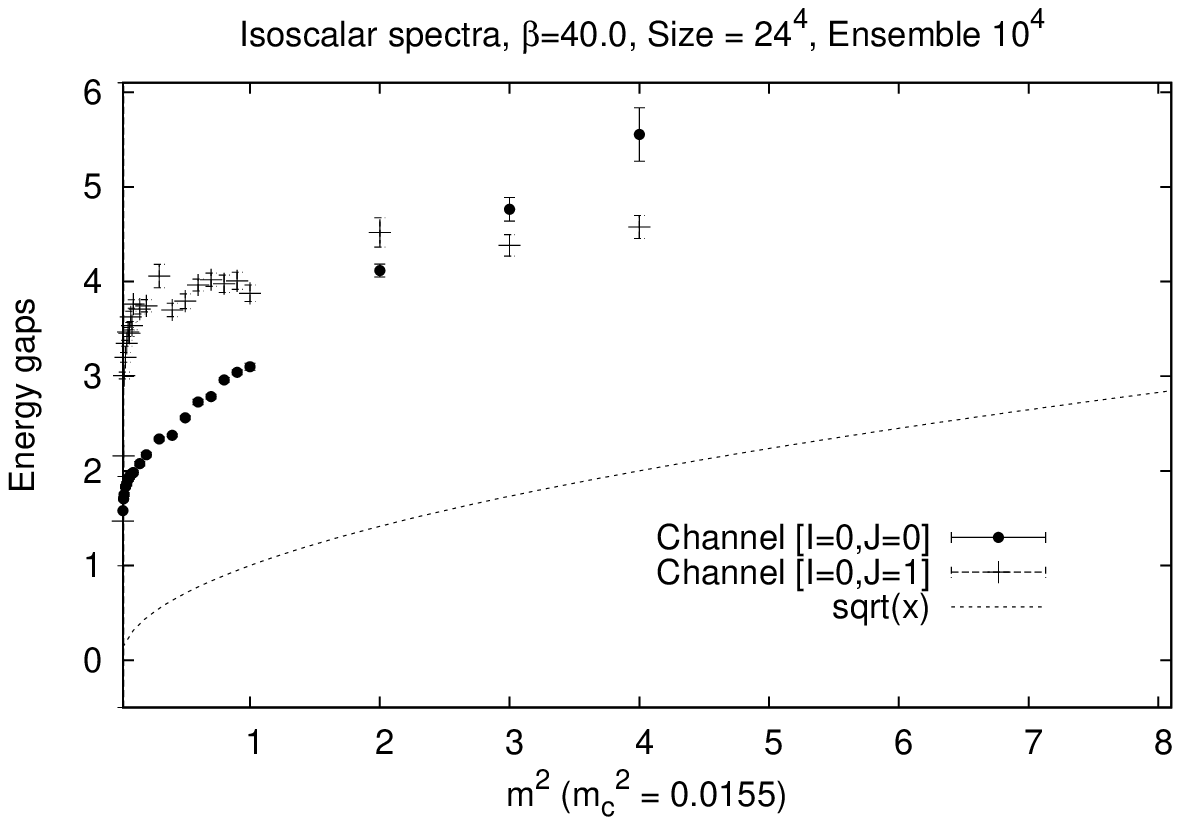}
\end{center}
\caption{Mass spectrum  ($m>>m_c$) of the isoscalars 
for $\beta=40$.}
\label{fig.7}
\end{figure}

The Figures \ref{fig.2.0}, \ref{fig.2},  \ref{fig.3} and  \ref{fig.4}
illustrate the fact that for $m^2>>m_{_{\rm TL}}^2$ gauge
vector mesons are present in the spectrum of the
lattice Massive Yang-Mills theory (\ref{lat.2}). 
The mass ($\Delta_1$) follows roughly the bare value $m$ and
looks not to depend much on $\beta$. 
The fit shown in the figures is performed by using the function
\begin{eqnarray}
\sqrt {m^2}\Big[1+ (A\ln m^2 +B)\Big]
\label{sim.4.4}
\end{eqnarray}
inspired by the expression of the self-energy in perturbation theory.
It parameterizes the departure of the gap from the bare value $m$.
\newpage
\clearpage
\newpage
The scalar isospin ($I=0$) states have  spin zero ($J=0$)
and spin ($J=1$). The Figs. \ref{fig.5.0}, \ref{fig.5}, \ref{fig.6} 
and \ref{fig.7} show the
numerical results. The patterns are not as clear as in the case
$I=1$. The energy gaps are much larger than $m$.
These states are not present in the naive continuum limit of zero
spacing, if perturbation theory is used. 
\par
Now we discuss the region where $m_c\le m<1$. 
For large $\beta$ ($\beta=3,10,40$)
two exponentials (see eq. (\ref{sim.4})) are necessary in the
region $m^2\sim m_c^2 $ in order to fit the time correlators
(\ref{sim.2}). For $\beta=1.5$ a single exponential fit works
well also for  values of $m\sim m_{_{\rm TL}}$ where the inflection points show up.
\par
\begin{figure}[hp]
\begin{center}
\includegraphics[clip,width=1\textwidth,,height=0.4\textheight]
{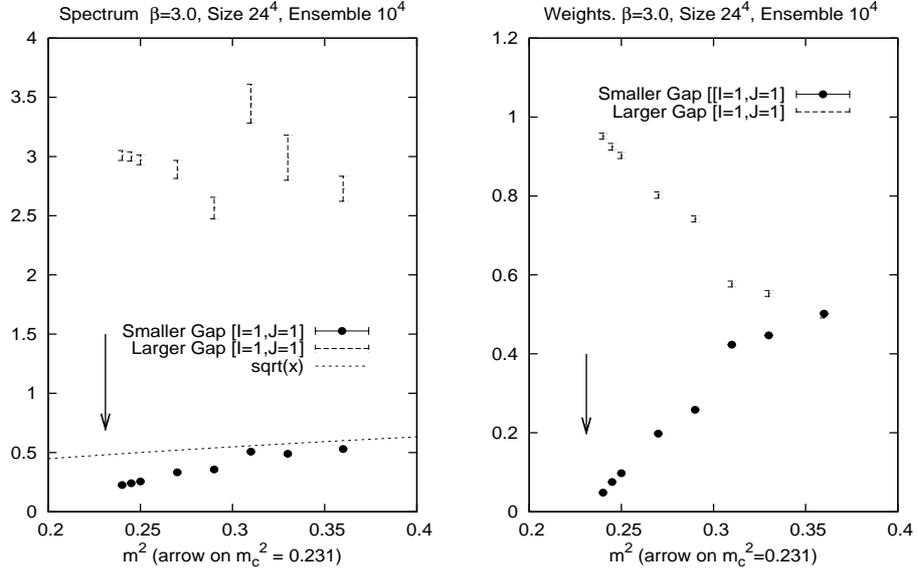}
\end{center}
\caption{Mass spectrum  and weights ($m_c\le m $) of the isovector
for $\beta=3$.}
\label{fig.8}
\end{figure}
\begin{figure}[hp]
\begin{center}
\includegraphics[clip,width=1\textwidth,,height=0.4\textheight]
{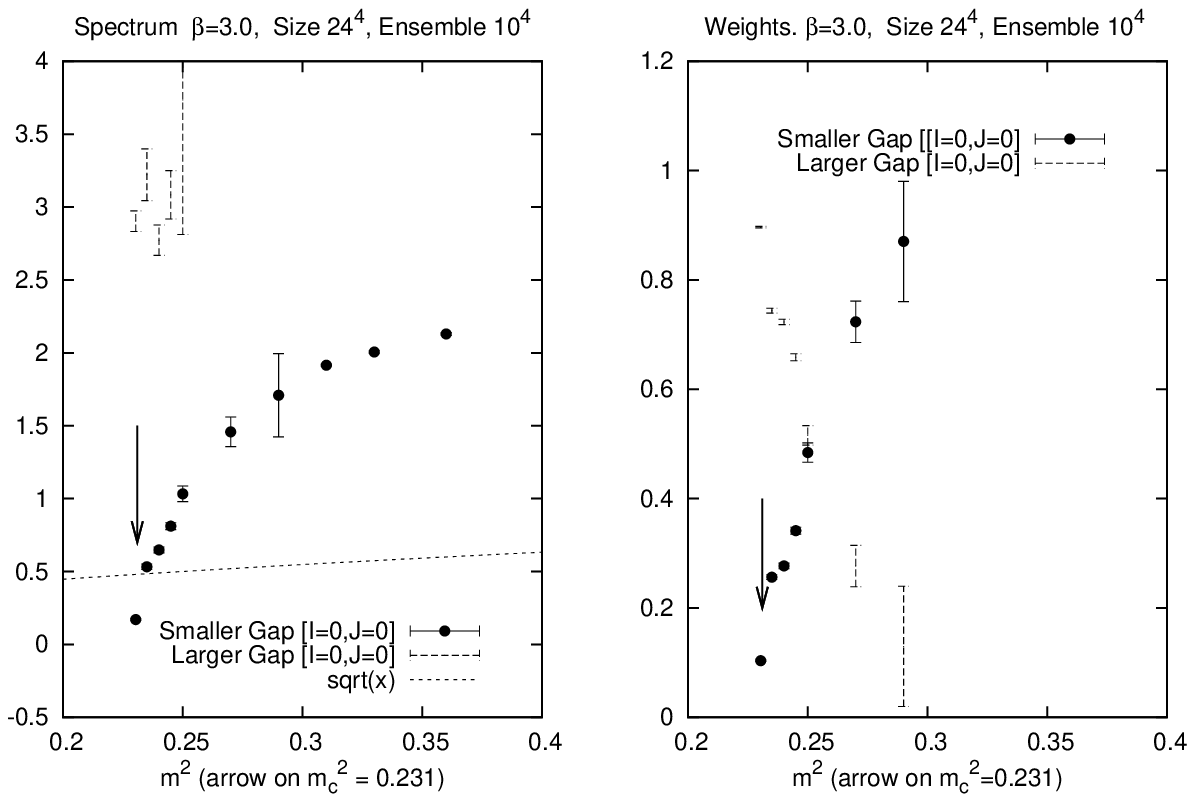}
\end{center}
\caption{Mass spectrum  and weights  ($m_c\le m $) of the $I=0,J=0$ 
for $\beta=3$.}
\label{fig.10}
\end{figure}
\begin{figure}[th]
\begin{center}
\includegraphics[clip,width=1\textwidth,,height=0.4\textheight]
{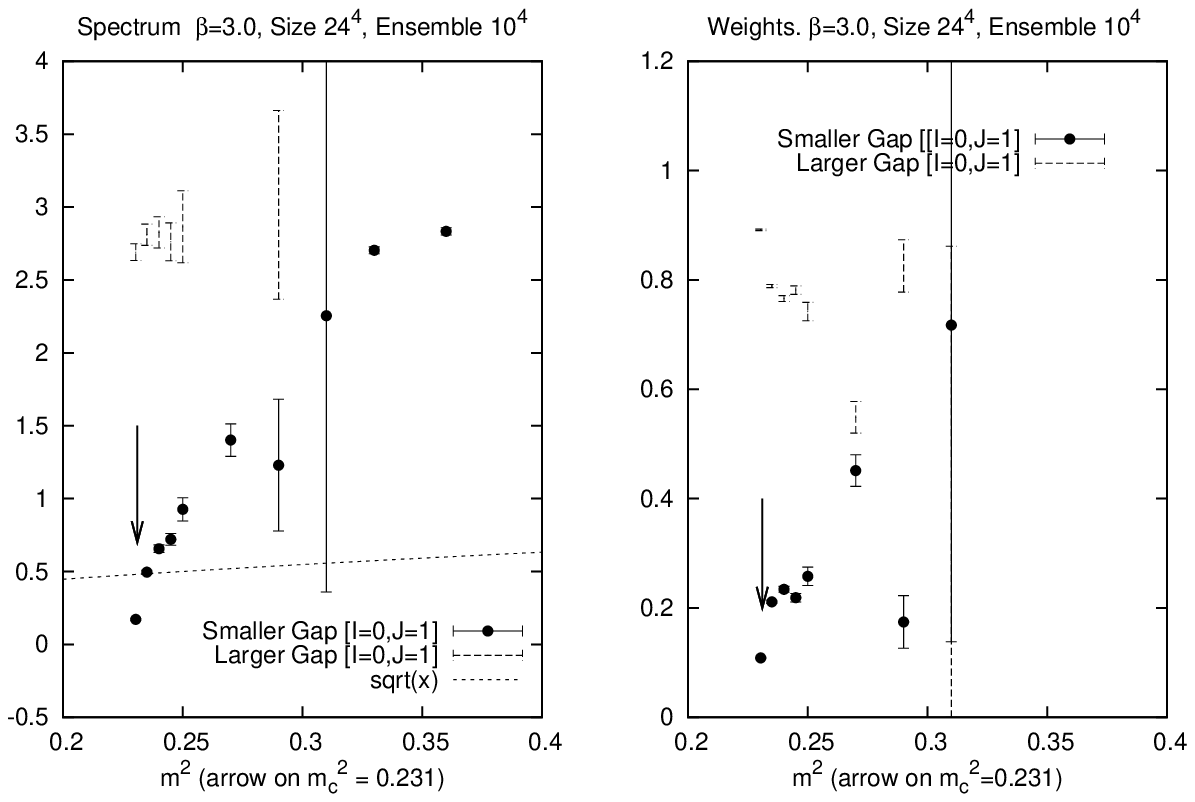}
\end{center}
\caption{Mass spectrum and weights  ($m_c\le m $) of the $I=0,J=1$
for $\beta=3$.}
\label{fig.11}
\end{figure}

The value of $m^2$ where the bifurcation occurs
is signaled by  sudden and very large errors on $\Delta_1$ and
 $\Delta_2$. The lower energy gap $\Delta_1$ follows the bare
value $m$, the weight $b_1$ becomes smaller and smaller 
by approaching $m_c$,
while at the same time $\Delta_2$ and $b_2$ increase.  The two pictures
in Fig. \ref{fig.8} show the bifurcation for $\beta=3$ in the isovector
channel. The pictures in Figs. \ref{fig.10} and \ref{fig.11} show a
similar phenomenon for  $\beta=3$ in the isoscalar channels ($J=0,1$). 
Our interpretation is that the lower energy gap is responsible
for the long range correlation, signaling the {\sl near}
fixed point at $\beta \to \infty$. The larger energy gap is
associated to the confining mechanism intervening in the
crossing of the TL.
A set of figures tries to illustrate these facts. In the isovector
channel ($J=1$) the pattern is very clear. By $m$ approaching
$m_c$ the lower gap $\sim m$ has a vanishing weight, while
the higher gap ($\gg m$) becomes dominant.
\par
A similar phenomenon occurs in the channels $I=0, J=0$ (see Fig. \ref{fig.10})
and $I=0, J=1$ (see Fig. \ref{fig.11}); 
however the onset of bifurcation is for lower
$m$ values and the patterns are not as clear as in the isovector
case.
\clearpage
\comment{
\subsection{Comment on the $\beta=1.5$ results}
In the case of $\beta=1.5$ the bifurcation of the energy
gap does not occur. This confirm that for $\beta < \beta_e$
the transition to screened/confined phase is smooth. The fit 
to the correlator data can be obtained by using only one
exponential in eq. (\ref{sim.4}). As shown in Fig. \ref{fig.2.0}
the value of the gap starts departing from the bare value
$m$ when the TL is approached. Then it has a steady increasing
behavior. For $\beta > \beta_e$ after the onset of the bifurcation
the lower gap has a steady decreasing weight and the higher gap
is replacing it. 
}
\par
We have repeated this analysis  for $\beta=10$ 
\comment{
in Figs. \ref{fig.8.10},\ref{fig.10.10} and  \ref{fig.11.10} 
}
and for $\beta=40$. The features are very similar to the
case $\beta=3$, thus we shall not provide further pictures
to illustrate the bifurcation phenomena for these cases.
\comment{
in Figs. \ref{fig.8.40}, \ref{fig.10.40} and in \ref{fig.11.40}.
}
\subsection{Comments on the Spectrum}
We summarize the comments on the spectrum resulting
from the lattice simulations. There is some non trivial time
correlation in the two-point function for the channels with quantum
numbers $(I=1,J=1)$, $(I=0,J=0)$ and $(I=0,J=1)$.
In the channel $I=1$ and $J=0$ we find zero correlation
for $t>0$ in eq. (\ref{sim.2}). No fit  of the function in eq. 
(\ref{sim.4}) is provided for this channel.
\par
The departure from the
bare value $m$ of the energy gap for $m>>1$ is common
to the values of $\beta=1.5,3, 10, 40$. 
\par
By approaching $m\sim m_c$ a single exponential
fitting of the time correlators is inadequate. A linear
combination of two exponentials provides a very good fit. 
Thus at some
value of $m$ (depending on $\beta$), the single gap
bifurcates: one follows the $m$ line while the other
is much larger. Moreover the weight of the lower vanishes
for $m\to m_c$.  This phenomenon is most evident in the
channel $I=1, J=1$. In the other channels the onset
of the bifurcation is faint and at smaller values
of $m$. Our scenario is the following: in approaching the
TL the lower gap provides the large correlation length,
while the large gap is the manifestation of the 
screening/confining mechanism, which becomes dominant
for $m\sim m_c$. 
\par
For $m\sim m_c$ there is some drastic changes in the 
energy gaps of the isoscalar channels (for both $J=0,1$):
one notices a sharp drop, even below the bare value $m$.
This fact sustains the scenario where 
bound states arise in the isoscalar channels because (i)
the gap energy becomes lower than the threshold and (ii) the
vector mesons decouple (very small weight in the two-point
functions). 
%
\section*{Acknowledgements}
%
We  gratefully acknowledge the warm hospitality of the
Center for Theoretical Physics at MIT, Massachusetts,
where part of the present work has been done.
We profited of many stimulating discussions with the
colleagues at the Department of Physics of the University
of Pisa.
\comment{
\newpage
\par
The numerical results are presented in the Tables \ref{tab:b3_isovector},
\ref{tab:b3_isoscalar_scalar} and \ref{tab:b3_isoscalar_vector}
for $\beta=3$ and ($I=1,J=1$),  ($I=0,J=0$) and  ($I=0,J=1$).
Similarly Tables 
\ref{tab:b10_isovector},
\ref{tab:b10_isoscalar_scalar} and \ref{tab:b10_isoscalar_vector}
list the results for $\beta=10$ and Tables 
\ref{tab:b40_isovector},\ref{tab:b40_isovector_continued},
\ref{tab:b40_isoscalar_scalar} and \ref{tab:b40_isoscalar_vector}
for $\beta=40$. 
The tables list $m^2$, $\Delta_1$, $b1$, $\Delta_2$, $b_2$  
(ordered according to $\Delta_1< \Delta_2$) and fit-$\chi^2$. 
Statistical errors are displayed. 
}%

\clearpage 
\end{document}